# Solving all laminar flows around airfoils all-at-once using a parametric neural network solver

Wenbo Cao[a,b,c], Shixiang Tang[c], Qianhong Ma[c,d], Wanli Ouyang[c], Weiwei Zhang[a,b,e],*

[a] *School of Aeronautics, Northwestern Polytechnical University, Xi'an 710072, China*
[b] *International Joint Institute of Artificial Intelligence on Fluid Mechanics, Northwestern Polytechnical University, Xi'an, 710072, China*
[c] *Shanghai Artificial Intelligence Laboratory, Shanghai, 200232, China*
[d] *School of Mathematical Sciences, Shanghai Jiao Tong University, Shanghai, 200240, China*
[e] *National Key Laboratory of Aircraft Configuration Design, Xi'an 710072, China*
* Corresponding author. E-mail address: aeroelastic@nwpu.edu.cn

**Abstract.** Recent years have witnessed increasing research interests of physics-informed neural networks (PINNs) in solving forward, inverse, and parametric problems governed by partial differential equations (PDEs). Despite their promise, PINNs still face significant challenges in many scenarios due to ill-conditioning. Time-stepping-oriented neural network (TSONN) addresses this by reformulating the ill-conditioned optimization problem into a series of well-conditioned sub-problems, greatly improving its ability to handle complex scenarios. This paper presents a new solver for laminar flow around airfoils based on TSONN and mesh transformation, validated across various test cases. Specifically, the solver achieves mean relative errors of approximately 3.6% for lift coefficients and 1.4% for drag coefficients. Furthermore, this paper extends the solver to parametric problems involving flow conditions and airfoil shapes, covering nearly all laminar flow scenarios in engineering. The shape parameter space is defined as the union of 30% perturbations applied to each airfoil in the UIUC airfoil database, with Reynolds numbers ranging from 100 to 5000 and angles of attack spanning from -5° to 15°. The parametric solver solves all laminar flows within the parameter space in just 4.6 day, at approximately 40 times the computational cost of solving a single flow. The model training involves hundreds of millions of flow conditions and airfoil shapes, ultimately yielding a surrogate model with strong generalization capability that does not require labeled data. Specifically, the surrogate model achieves average errors of 4.6% for lift coefficients and 1.1% for drag coefficients, demonstrating its potential for high generalizability, cost-effectiveness, and efficiency in addressing high-dimensional parametric problems and surrogate modeling.

# 1 Introduction

Traditional numerical methods for partial differential equations, such as finite difference, finite volume, and finite element methods, have been indispensable tools in computational fluid dynamics (CFD) for decades. They offer high-fidelity solutions to complex flow problems, driving advancements in engineering design, climate modeling, and fundamental fluid mechanics. However, as scientific research and engineering applications advance, the computational demands of certain tasks have grown exponentially. Applications such as optimal control, aerodynamic shape optimization, and real-time flow field prediction often require repeated evaluations of fluid dynamics over a wide range of scenarios, making traditional numerical approaches computationally prohibitive.

With advances in deep learning and computational infrastructure, data-driven modeling has emerged as a powerful approach to overcoming these challenges, enabling rapid predictions of complex flow fields. Leveraging various network architectures [1-7], these methods have been applied across a wide range of scenarios [8], yielding impactful results and showcasing substantial potential for engineering applications. Despite their advantages, data-driven approaches often heavily rely on expensive training data, which are often computed using expensive traditional numerical methods [9]. Consequently, the limited number of conditions represented in the training data in practice poses significant challenges to the accuracy and generalization capability of the data-driven surrogate models.

Innovative techniques such as physics-informed neural networks (PINNs) [10], the deep Ritz method [11], and the deep Galerkin method [12] are changing the way to solve forward and inverse PDE problems. These methods optimize the parameters of the neural network to simultaneously minimize the losses associated with the PDE, boundary conditions, and initial conditions, allowing the network's output to effectively approximate the solution of the PDE system. Numerous PINN-like methods have emerged in recent years, yielding impressive results across various challenges in computational science and engineering [13-16]. As a neural network-based mesh-free approach, PINNs offer significant advantages in mitigating the curse of dimensionality, achieving remarkable success in solving high-dimensional PDEs and parametric PDEs, and emerging as a promising tool for surrogate modeling. Sirignano et al. [12] proposed the Deep Galerkin Method (DGM) for solving PDEs in up to 200 dimensions and solving parametric Burgers' equations. Sun et al. [17] applied PINNs to solve parametric Navier-Stokes equations involving viscosity or geometry. Wang et al. [18] combined PINNs with DeepONet to solve parametric PDEs related to initial conditions, boundary conditions, or inputs across a range of fundamental equations in various fields, a method referred to as operator learning. For the flow around airfoils, Sun et al. [19] used PINNs to solve parametric Navier-Stokes equations at low Reynolds numbers with respect to geometry, demonstrating the potential of the model in shape design optimization. Cao et al. [20] combined PINNs with mesh transformation techniques to solve parametric Euler equations in inviscid flow. Their model incorporated all state parameters of interest for inviscid flow, including flow conditions and shape, and introduced a pretraining and fine-tuning approach to enhance surrogate model performance on specific tasks.

Despite their promising potential, these methods remain largely inapplicable for parametric solutions of laminar flows due to the ill-conditioning challenges of PINNs. In our previous work [21], we proposed the time-stepping-oriented neural network (TSONN), which transforms the ill-conditioned problem of PINNs into a sequence of well-conditioned sub-optimization problems. This method has been successfully applied to solve lid-driven cavity flow, two-dimensional laminar flow around an airfoil, and three-dimensional laminar flow around a wing, all at a Reynolds number of 5000. It has also recently been extended to solve high-Reynolds-number wall-bounded turbulence around airfoils governed by Reynolds-averaged Navier-Stokes equations coupled with the Spalart-Allmaras turbulence model [22].

In this work, we significantly extend TSONN to address the parametric problem of laminar flow around airfoils, aiming to develop an engineering-ready parametric solver and build a surrogate model with robust generalization capabilities. The main contributions of our work can be summarized as follows:

1. We develop a laminar flow solver for airfoils based on TSONN and mesh transformation and validate the solver under a wide range of flow conditions and shapes.

2. We solve a parametric problem covering various laminar flow scenarios in engineering, resulting in a surrogate model with strong generalization capability, highlighting the advantages of TSONN in solving high-dimensional parametric problems.

3. We verify that both high-dimensional and low-dimensional airfoil representation methods achieve similar convergence curves and results, demonstrating the solver's ability to automatically extract low-dimensional features.

The remainder of this paper is organized as follows. In Section 2, we present the problem setup, and the various techniques employed, followed by a detailed introduction to the parametric solver. Section 3 discusses the results of TSONN in solving both single-flow and parametric problems. Finally, Section 4 provides a summary and outlines potential directions for future research.

## 2 Methodology

### 2.1 Problem setting

We consider the two-dimensional incompressible Navier-Stokes equations for solving viscous flows around airfoils. The governing equations are expressed as follows:

$$\begin{aligned} \boldsymbol{V} \cdot \nabla \boldsymbol{V} + \nabla p - \Delta \boldsymbol{V} / Re &= 0 \\ \nabla \cdot \boldsymbol{V} &= 0 \end{aligned} \quad (1)$$

where $\boldsymbol{V} = [u, v]$ is the velocity vector; $p$ is the pressure; and $Re$ is Reynolds number. The schematic of the computational domain, mesh, and boundary conditions for solving the flow around airfoils is shown in Figure 1. The flow satisfies the no-slip boundary condition on the wall with $u = 0, v = 0$. At the velocity inlet, the conditions are $u = \cos(\alpha), v = \sin(\alpha)$, where $\alpha$ is the angle of attack. At the pressure outlet, the condition is $p = 0$.

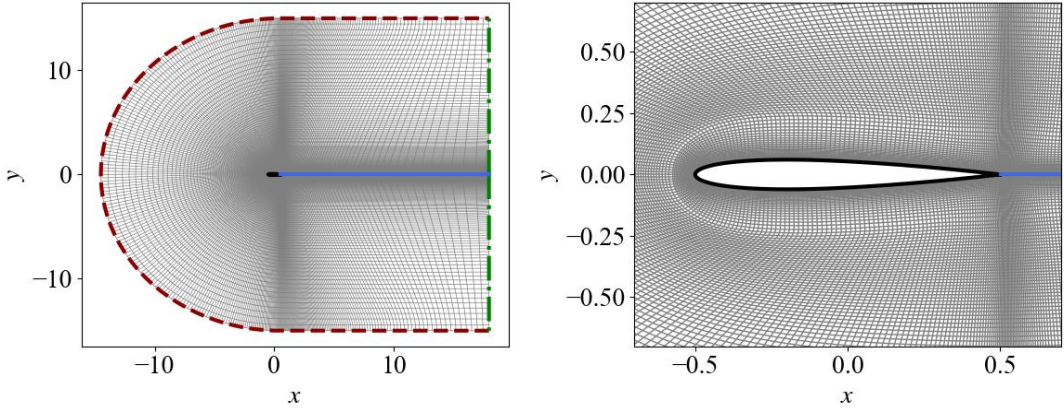

Figure 1. Schematic illustration of the computational domain, mesh, and boundary conditions. The red line indicates the velocity inlet boundary, the green line indicates the pressure outlet boundary, the black line indicates the wall boundary, and the blue line represents an artificial overlapping boundary caused by mesh transformation.

The flow around airfoils focuses on the analysis of flow under given flow conditions and airfoil shapes, with particular attention to the wall pressure coefficient distribution ($C_p$) and wall skin friction coefficient distribution ($C_f$) (Equation (2)), as well as the lift coefficient ($C_l$) and drag coefficient ($C_d$) obtained by integrating these distributions over the surface. For more details, please refer to reference [23].

$$C_p = (p - p_\infty)/(0.5\rho_\infty V_\infty^2) = 2p$$
$$C_f = \tau_w /(0.5\rho_\infty V_\infty^2) = \frac{2}{Re}\left(\frac{\partial u_{\tan}}{\partial n}\right)_w \cdot \quad (2)$$

2.2 Physics-informed neural networks and time-stepping-oriented neural network

In this section, we briefly introduce PINNs and TSONN. A typical PINN employs a fully connected deep neural network (DNN) architecture to represent the solution $q$ of the dynamical system. The network takes the spatial state $x \in \Omega$ and temporal state $t \in [0,T]$ as the input and outputs the approximate solution $\hat{q}(x,t;\boldsymbol{\theta})$. The spatial domain typically has 1-, 2-, or 3-dimensions in most physical problems, and the temporal domain may be nonexistent for time-independent (steady) problems. The result of PINNs is determined by the network parameters $\boldsymbol{\theta}$, which are optimized with respect to loss function during the training process.

In, we formulate the loss function as the mean squared error (MSE) of the residual vector, following the approach in reference [21]. The residual vector $\boldsymbol{f}(\boldsymbol{q})$ encompasses the PDE residual $\boldsymbol{g}(\boldsymbol{q})$, the boundary condition residual $\boldsymbol{h}(\boldsymbol{q})$, and initial condition residual $\boldsymbol{i}(\boldsymbol{q})$ (if any), requiring a trade-off between different components through appropriate relative weights $\lambda_{PDE}$, $\lambda_{BC}$, and $\lambda_{IC}$, expressed as follows:

$$\boldsymbol{f}(\boldsymbol{q}) = \begin{bmatrix} \lambda_{PDE}\boldsymbol{g}(\boldsymbol{q}) \\ \lambda_{BC}\sqrt{N_g/N_h}\,\boldsymbol{h}(\boldsymbol{q}) \\ \lambda_{IC}\sqrt{N_g/N_i}\,\boldsymbol{i}(\boldsymbol{q}) \end{bmatrix} = 0 \quad (3)$$

where $N_g$, $N_h$ and $N_i$ are the dimension of PDE residual, boundary condition residual, and initial condition residual respectively. One calculates the residual vector $\boldsymbol{f}(\boldsymbol{q})$ over a series of $m$ collocation points $D = \{x_i,t_i\}_{i=1}^m$ by automatic differentiation [24], and then minimizes the loss

$$\mathcal{L} = \frac{1}{N}\left\|\boldsymbol{f}\left(\boldsymbol{q}(\cdot;\boldsymbol{\theta})\right)\right\|^2 \quad (4)$$

In many scenarios, achieving stable training and accurate results with PINNs remains a challenge due to their ill-conditioning. In our previous work [21], we proposed TSONN, which decomposes the original equations into a sequence of implicit pseudo time-stepping equations (as shown in Equation (5)), thereby transforming the ill-conditioned optimization problem into a sequence of well-conditioned sub-optimization problems.

$$\begin{aligned} &PINNs: f(q) = 0 \\ &TSONN: f(q) - (q - q_n)/\Delta\tau = 0, q_n = \hat{q}(\cdot;\theta_n), n = 0,1,\cdots,N \end{aligned} \quad (5)$$

For supervised boundary conditions, $q_n$ in Equation (5) is replaced by target values. Algorithm 1 provides an implementation for TSONN. For more detailed information about TSONN, please refer to [21, 25].

**Algorithm 1:** Time-stepping-oriented neural network (TSONN)
**Input:** Initial $\theta$, outer iterations $N$, inner iterations $K$, pseudo time step $\Delta\tau$.
1: for $n = 1, 2, \cdots, N$ do
2:   (a). Randomly sample collocation points and initialize an LBFGS optimizer with $K$ iterations per optimization step.
3:   (b). Get the state $q_n = q(\cdot;\theta)$
4:   (c). Perform an optimization step using the optimizer to minimize the loss
$$\mathcal{L}(\theta) = \frac{1}{N}\|f(q(\cdot;\theta)) - (q(\cdot;\theta) - q_n)/\Delta\tau\|^2$$
5: end
**Output:** $\hat{q}(\cdot;\theta)$

2.3 Mesh transformation and volume-weighted technique

Mesh transformation is one of the essential techniques in finite difference methods, employed to convert non-uniform grids into uniform rectangular grids with mappings $\xi = \xi(x, y)$ and $\eta = \eta(x, y)$. For example, the grid shown in Figure 1 can be cut at the blue position and stretched into a uniform rectangular grid. Previous studies [22, 26] combined PINNs and TSONN with mesh transformations to solve inviscid flows and high-Reynolds-number viscous flows around airfoils. These approaches leveraged neural networks to learn flow in the uniform computational space rather than in the physical space. Mesh transformation serves to scale up the regions near the airfoil, thereby aiding the neural network in learning the flow.

When using mesh transformation, the first-order derivatives of $q$ in the physical space are evaluated by the transformation

$$\begin{aligned} q_x &= q_\xi \xi_x + q_\eta \eta_x \\ q_y &= q_\xi \xi_y + q_\eta \eta_y \end{aligned} \quad (6)$$

where

$$\begin{bmatrix} \xi_x & \eta_x \\ \xi_y & \eta_y \end{bmatrix} = \begin{bmatrix} x_\xi & y_\xi \\ x_\eta & y_\eta \end{bmatrix}^{-1} = J^{-1} \quad (7)$$

The terms $q_\xi$ and $q_\eta$ are evaluated using automatic differentiation. The terms $x_\xi, y_\xi, x_\eta$ and $y_\eta$ are called metrics, which are obtained through finite differences of the grid. For simplicity, only the calculation of first-order derivatives is provided. For details on the computation of second-order derivatives, please refer to reference [22].

When using grid transformation, the neural network is employed to learn the flow in the computational space. As a result, the computational domain no longer deforms with the shape changes; instead, the shape is implicitly embedded in the metric coefficients. Therefore, the parametric problem with respect to the shape and computational domain is transformed into the parametric problem with respect to the metric coefficients in the equations, which is particularly advantageous for the implementation and solution of parametric problem.

In addition, we employ a volume-weighted PDE residual [27], which is used to scale the weights of grid points that are very dense near the wall. The technique has been validated to perform better in non-uniform mesh distributions for many problems [20, 21, 27]. Consequently, $g(\boldsymbol{q})$ is redefined as:

$$g(\boldsymbol{q}) \leftarrow \frac{g(\boldsymbol{q})\boldsymbol{v}}{\|\boldsymbol{v}\|_2 / \sqrt{N_g}} \tag{8}$$

where the vector $\boldsymbol{v}$ encompasses the mesh volume at each mesh point, and the denominator in Equation (8) normalizes the influence of $\boldsymbol{v}$.

2.4 A parametric solver for laminar flow around airfoils

To obtain continuous solutions within a given parameter space, we include all state parameters of interest in the flow around airfoils in the model input. As shown in Figure 2, the input of the parametric solver encompasses all conditions in viscous flow around airfoils, including the coordinate in computational space, Reynolds number $Re$, angle of attack $\alpha$, and shape parameter vector $\boldsymbol{s}$. We consider two approaches for shape representation. The first uses 6 Class-Shape Transform (CST) [28] parameters for each of the upper and lower surfaces, resulting in a 16-dimensional high-dimensional parametric problem. The second directly represents the shape using the *y*-coordinates of 200 discrete points, with identical *x*-coordinates for all shapes, leading to a high-dimensional parametric problem of up to 204 dimensions. The model outputs are the velocity components $u, v$ and the pressure $p$. Since the solver is based on mesh transformation, it requires using the shape parameter vector from the input to compute the grid and metrics. To obtain these terms efficiently, the structured mesh of NACA0012 as shown in Figure 1 is used as the base mesh, and the meshes of other airfoils are generated by mesh deformation based on the Radial Basis Function (RBF) interpolation [29].

When solving the parametric problem, we randomly select 30000 interior residual points and 1000 boundary points for each batch, along with randomly chosen flow conditions and shapes within the given parameter space. These are then concatenated into input vector. As a result, the flow conditions and shapes for each collocation point are different, which helps to include as many flow conditions and shapes as possible in each batch. By assigning different conditions to each collocation point and resampling the points in every outer iteration, we can account for hundreds of millions of flow conditions and shapes when solving the parametric problem. This enables the model to achieve strong generalization capability. When solving a single flow, we can directly obtain the current shape's mesh and metrics across all points based on mesh deformation technique, requiring only a one-time calculation. However, when solving the parametric problem with respect to shapes, it is impractical to compute the deformed mesh and metrics for every shape. Since each collocation point corresponds to one shape and one coordinate in computational space, we only calculate the deformed coordinates $x$, $y$ of the grid point and its neighbor girds and then calculate those terms by finite difference.

In all cases, the relative weights are $\lambda_{PDE}=100$, $\lambda_{BC}=1$, the inner iterations $K$ is 500, and the pseudo-time step is $\Delta\tau=0.3$. These hyperparameters have been validated in previous studies [21, 22]. In all cases of solving a single flow, the solver employs a fully connected DNN architecture with 6 hidden layers, each containing 128 neurons, equipped with the hyperbolic tangent activation functions (tanh). When solving the parametric problem in Section 3.3, we increase the number of hidden layers from 6 to 10 to enhance the model's representational capacity.

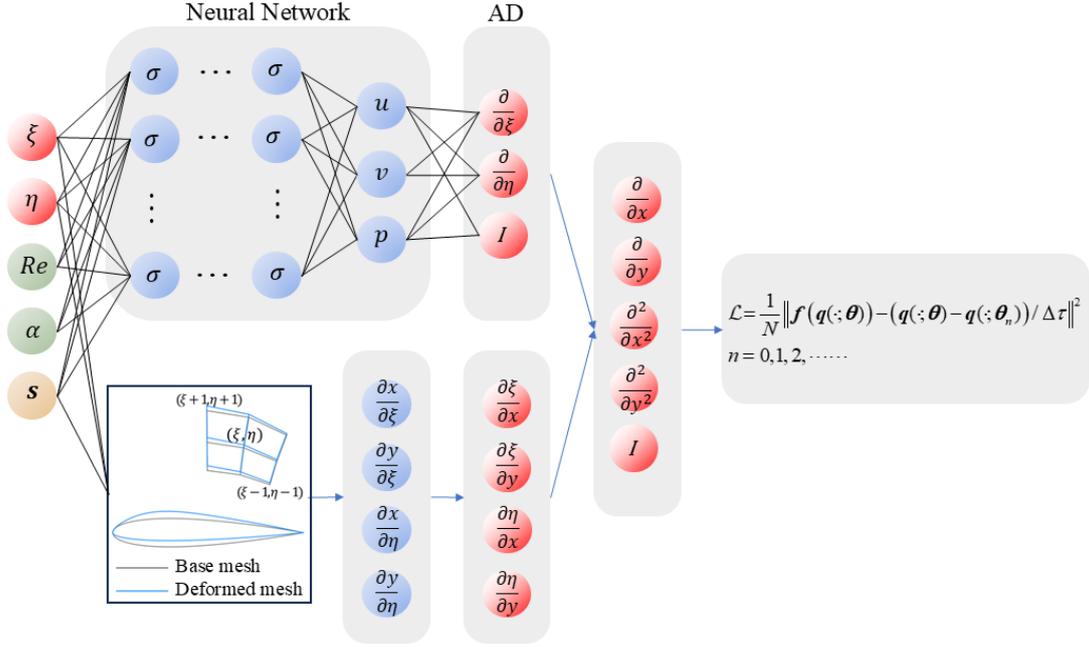

Figure 2. Schematic diagram of the parametric solver based on TSONN and mesh transformation.

## 3 Results

In this section, we validate the proposed parametric solver for laminar flow around airfoils across a wide range of test cases. We employ our in-house second-order finite volume method (FVM) solver to simulate flows and generate reference solutions. The solver utilizes the Roe scheme for computing inviscid fluxes, with interface values reconstructed using a second-order least-squares approach. Viscous fluxes are discretized with a standard second-order central scheme. The mesh used in the FVM is consistent with the one employed by the PINNs, as shown in Figure 1. All our model training is performed on an NVIDIA GeForce RTX 4090 GPU. The three surrogate models presented in Section 3.3 are publicly available at https://github.com/Cao-WenBo/ParametricSolverForAirfoilFlows.

3.1 Case configurations

Table 1 lists the cases used to validate the parametric solver. Their Reynolds numbers range from 100 to 5000, covering nearly the entire Reynolds number range for laminar flow around airfoils. The angle of attack ranges from -5 to 15 degrees, encompassing the angles of interest for the flow around airfoils in engineering applications. Several classic airfoils from the UIUC database, shown in Figure 3, are considered to validate the solver.

Table 1. Case configurations for evaluating the parametric solver.

|       | $Re$ | $\alpha$ | Airfoil  |
|-------|------|----------|----------|
| Case1 | 100  | 15°      | NACA0012 |
| Case2 | 200  | 11°      | NACA4412 |
| Case3 | 500  | 7°       | RAE2822  |
| Case4 | 1000 | 3°       | RAE5214  |
| Case5 | 2500 | -1°      | S2050    |
| Case6 | 5000 | -5°      | S9000    |

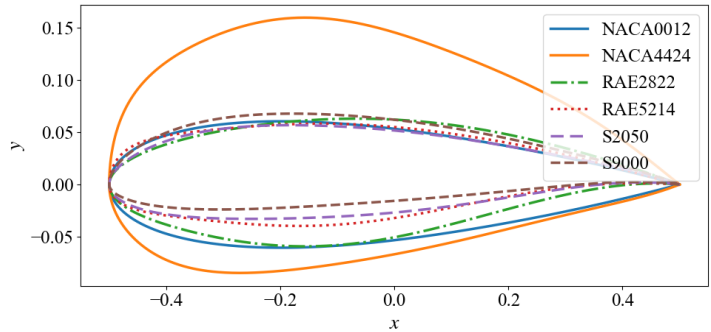

Figure 3. Airfoils for evaluating the parametric solver.

3.2 Solving single flows under various conditions

We first validate the solver's capability in solving a single flow and compare its results with those obtained by PINNs. To obtain the results for PINNs, we set the reciprocal of the pseudo-time step $1/\Delta\tau$ in TSONN to 0 while keeping all other settings unchanged. The simulation results are shown in Figure 4 and Figure 5. We observe that the wall pressure coefficient and skin friction distributions obtained by TSONN are consistent with the reference solution, whereas PINNs exhibit significant errors, especially for the highest Reynolds number shown in Case 6. Figure 6 presents the loss and error convergence curves for Case 6 obtained by both TSONN and PINNs. We observe that although PINNs achieve more stable convergence of the loss function compared to TSONN, its error stagnates after a rapid initial decrease, underscoring the issue of ill-conditioning, which is consistent with previous observations regarding the lid-driven flow [25]. In contrast, TSONN, by alleviating ill-conditioning, achieves stable convergence in both the loss and error.

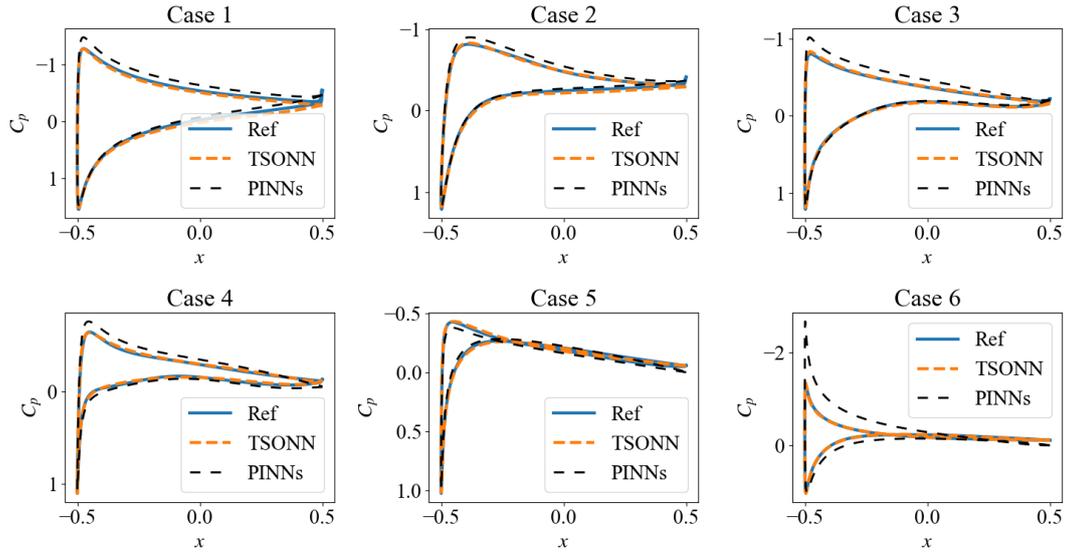

Figure 4. The wall pressure coefficient distributions for different cases obtained by TSONN and PINNs.

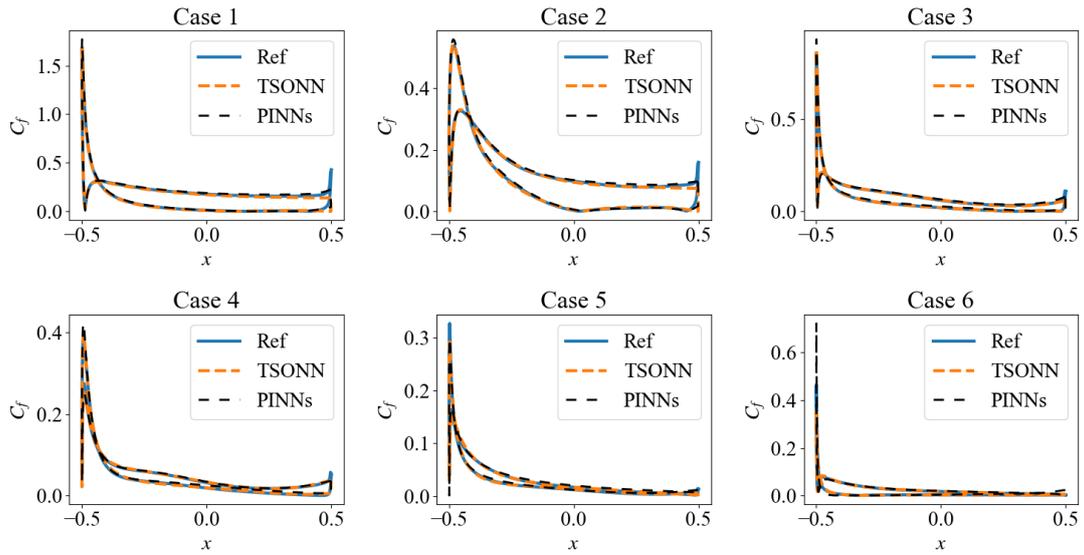

Figure 5. The wall skin friction coefficient distributions for different cases obtained by TSONN and PINNs.

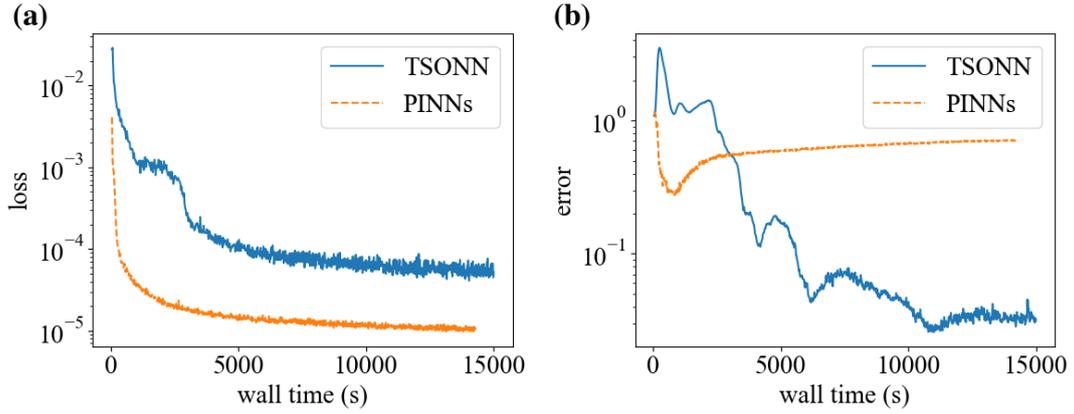

Figure 6. The convergence curves of (a) the loss function and (b) the relative $L_1$ error of the wall pressure coefficient distribution for Case 6, as obtained by TSONN and PINNs.

Table 2 provides a detailed listing of the lift coefficient and drag coefficient obtained by integrating the predicted the wall pressure coefficient distribution and wall skin friction coefficient distribution using the TSONN method, which is used to quantify the accuracy of the solver. The solver demonstrates satisfactory accuracy, with a mean relative error of 3.6% for the lift coefficient and 1.4% for the drag coefficient. As previously stated, the same hyperparameters are used across all cases. Therefore, these results demonstrate that the solver can robustly solve laminar flow around airfoils.

Table 2. Results and average errors obtained by TSONN for different cases. Note: Case 5 is excluded when calculating the mean relative error related to lift, as its excessively small value would result in disproportionately large and biased relative errors. The excluded results are highlighted in bold.

|  |  | $C_l$ | $C_{d,p}$ | $C_{d,f}$ | $C_d$ |
|---|---|---|---|---|---|
| Case 1 | Ref | 0.6626 | 0.2240 | 0.2796 | 0.5036 |
|  | Pred | 0.6727 | 0.2220 | 0.2607 | 0.4826 |
| Case 2 | Ref | 0.3223 | 0.1723 | 0.2099 | 0.3821 |
|  | Pred | 0.3498 | 0.1725 | 0.2045 | 0.3770 |
| Case 3 | Ref | 0.3581 | 0.0629 | 0.1220 | 0.1849 |
|  | Pred | 0.3741 | 0.0630 | 0.1203 | 0.1833 |
| Case 4 | Ref | 0.2191 | 0.0332 | 0.0844 | 0.1176 |
|  | Pred | 0.2222 | 0.0331 | 0.0856 | 0.1187 |
| Case 5 | Ref | **-0.0441** | 0.0175 | 0.0517 | 0.0692 |
|  | Pred | **-0.0601** | 0.0172 | 0.0519 | 0.0691 |
| Case 6 | Ref | -0.1292 | 0.0302 | 0.0246 | 0.0548 |

|  | Pred | -0.1321 | 0.0307 | 0.0247 | 0.0554 |
|---|---|---|---|---|---|
| Mean relative error | | 3.64% | 0.81% | 2.15% | 1.42% |
| Mean absolute error | | 0.0126 | 0.0005 | 0.0046 | 0.0049 |

3.3 Solving parametric problems

In section 3.2, we validate the solver's effectiveness in solving a single flow. Although the accuracy of the solver is satisfactory, its efficiency remained much slower than that of traditional numerical methods. In this section, we consider high-dimensional parametric problems that include nearly all laminar flow scenarios around airfoils encountered in engineering applications. The goal is to obtain a surrogate model with strong generalization capability. The target parameter space for Reynolds number and angle of attack is $Re \in [10, 5000]$, $\alpha \in [-5°, 15°]$. Similar to our previous study [20], the target parameter space of CST parameters is not a continuous interval. This is because a large interval may result in non-physical airfoils, while a small interval may lead to minor shape changes. Instead, the shape parameter space is the union of spaces generated by randomly perturbing each shape in the UIUC airfoil database, a rich and professional database containing approximately 1600 airfoils. Specifically, for each collocation point, we randomly select a shape from the UIUC database and add ±30% random perturbation to its CST parameters, using the perturbed CST parameters as input. These practices ensure that the parameter space of CST is large enough to include almost all airfoils that may be encountered in practices. The airfoils from the UIUC database that we used are shown in Figure 7. Although we employed two shape representation methods, CST parameters and *y*-coordinates, we did not directly perturb the *y*-coordinates to generate the shape parameter space when using the *y*-coordinate representation. Instead, we still perturbed the CST parameters, as directly perturbing the *y*-coordinates could result in highly non-smooth shapes. In other words, the shape parameter spaces for both representation methods are identical, which facilitates a fair comparison of model performance.

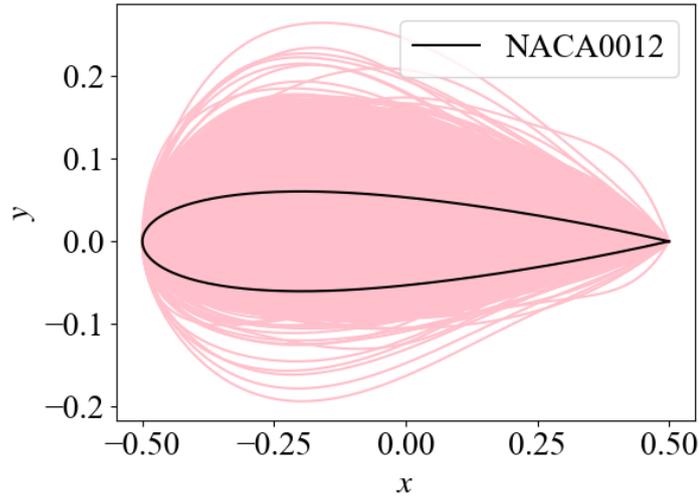

Figure 7. The airfoils from the UIUC database. When solving the parametric problem, each airfoil in the figure is randomly perturbed to generate a larger shape parameter space.

In the model training process, to avoid potential errors caused by insufficient sampling at the boundaries of the parameter space, we further expanded the actual sampling space of the state parameters, as shown in Table 1, Model 1 and Model 2 represent shapes using CST parameters and y-coordinates, respectively. To further demonstrate the capability of the current surrogate modeling approach to arbitrarily extend the model generalization space, we also consider an even larger parameter space, as shown by model 3.

|         | $Re$        | $\alpha$      | Shape representation | CST perturbation |
|---------|-------------|---------------|----------------------|------------------|
| Model 1 | (10,7000)   | (-8°,18°)     | CST                  | ± 35%            |
| Model 2 | (10,7000)   | (-8°,18°)     | $y$                  | ± 35%            |
| Model 3 | (1,10000)   | (-20°,20°)    | CST                  | ± 50%            |

Figure 8 presents the convergence history for solving the parametric problem and a single flow. We perform gradient descent involving 10,000 outer iterations, a total of $(30{,}000 + 1{,}000) * 10{,}000 = 3.1*10^8$ different collocation points. We observe that Model 1 and Model 2 exhibit similar convergence curves, indicating that with the same shape parameter space, using $y$-coordinates to represent airfoil shapes incurs a similar training cost as using CST parameters, despite resulting in a high-dimensional parametric problem with up to 204 dimensions. This suggests that the solver possesses the capability to automatically extract low-dimensional features. Moreover, using discrete coordinates to represent shapes offers potential advantages for handling more complex shapes in the future. When the parameter space is further expanded, the model still achieves stable convergence, although with slower convergence and lower accuracy compared to Model 1. Additionally, we find that solving the parametric problem requires approximately 40 times the computational cost of solving a single flow, highlighting the significant advantage of this solver for parametric problems.

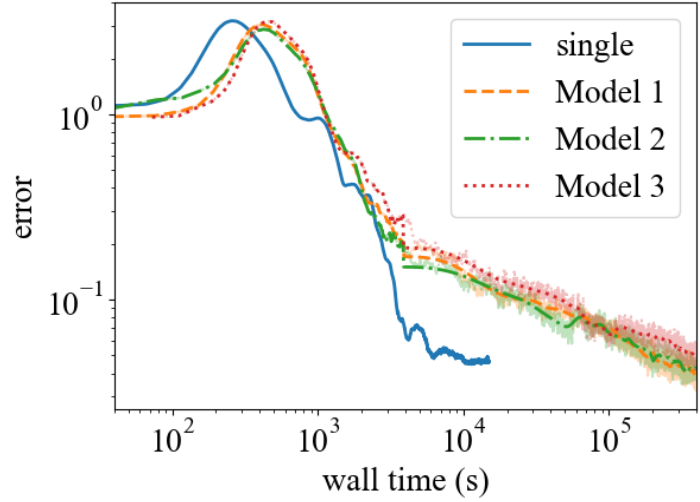

Figure 8. Error convergence curves for solving parametric problem and a single flow, where the error is the average relative $L_1$ error of the wall pressure coefficient distributions across the six cases listed in Table 1. The transparent lines represent the actual convergence curves, while the opaque lines represent Gaussian-filtered convergence curves for clearer visualization.

By solving the parametric problem, surrogate models are constructed, enabling near real-time prediction of any flow within the parameter space. Figure 9 and Figure 10 show the predicted wall pressure coefficient and skin-friction coefficient distributions for the six test cases listed in Table 1 by Model 1, respectively. The results demonstrate that the model maintains high consistency with the reference solutions. Table 3 provides a detailed comparison of lift and drag coefficient errors for each test case. The model achieves satisfactory accuracy, with average relative errors of 4.6% for lift coefficients and 1.12% for drag coefficients, comparable to the accuracy of solving a single flow. Additionally, Figure 11 and Figure 12 present the contour plots of flow fields for different cases, further illustrating the model's capability to predict flow fields effectively. These results indicate that even simple fully connected networks can achieve strong generalization capability.

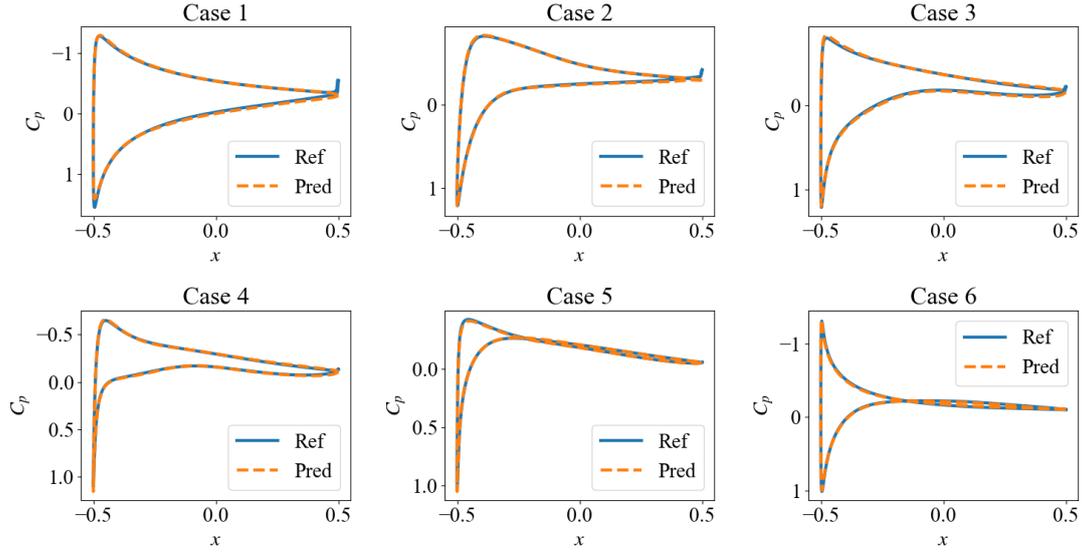

Figure 9. The wall pressure coefficient distributions predicted by Model 1 for Case 1 to 6.

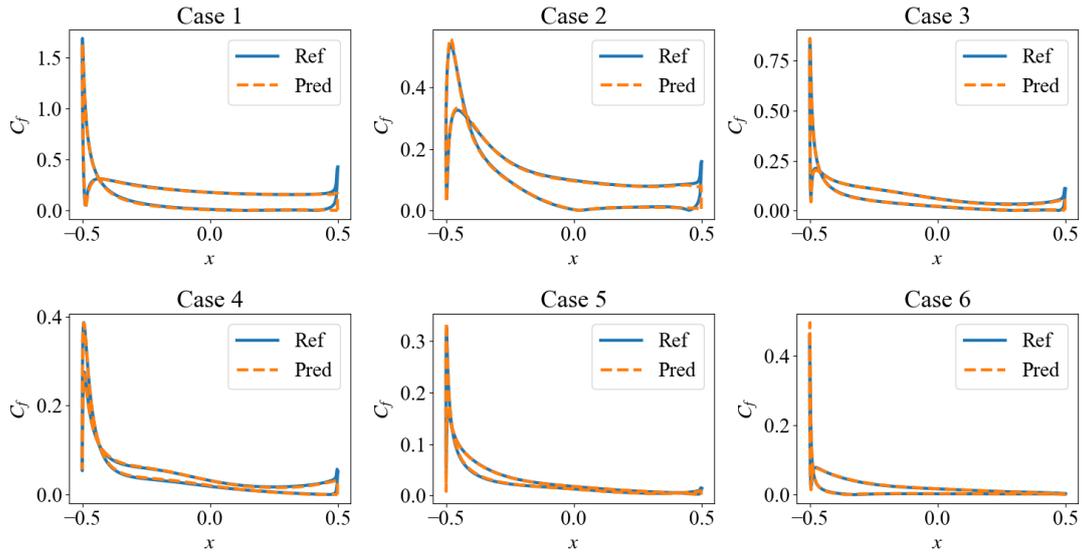

Figure 10. The wall skin friction coefficient distributions predicted by Model 1 for Case 1 to 6.

Table 3. Results and average errors predicted by Model 1. Note: Case 5 is excluded when calculating the mean relative error related to lift, as its excessively small value would result in disproportionately large and biased relative errors. The excluded results are highlighted in bold.

|  |  | $C_l$ | $C_{d,p}$ | $C_{d,f}$ | $C_d$ |
|---|---|---|---|---|---|
| Case 1 | Ref | 0.6626 | 0.2240 | 0.2796 | 0.5036 |
|  | Pred | 0.6860 | 0.2234 | 0.2659 | 0.4893 |
| Case 2 | Ref | 0.3223 | 0.1723 | 0.2099 | 0.3821 |
|  | Pred | 0.3340 | 0.1716 | 0.2061 | 0.3778 |
| Case 3 | Ref | 0.3581 | 0.0629 | 0.1220 | 0.1849 |
|  | Pred | 0.3793 | 0.0641 | 0.119 | 0.1831 |
| Case 4 | Ref | 0.2191 | 0.0332 | 0.0844 | 0.1176 |
|  | Pred | 0.2270 | 0.0338 | 0.0844 | 0.1182 |
| Case 5 | Ref | **-0.0441** | 0.0175 | 0.0517 | 0.0692 |
|  | Pred | **-0.0381** | 0.0177 | 0.0515 | 0.0692 |
| Case 6 | Ref | -0.1292 | 0.0302 | 0.0246 | 0.0548 |
|  | Pred | -0.1374 | 0.0312 | 0.0243 | 0.0555 |
| Mean relative error |  | 4.61% | 1.47% | 1.80% | 1.12% |
| Mean absolute error |  | 0.0131 | 0.0007 | 0.0035 | 0.0036 |

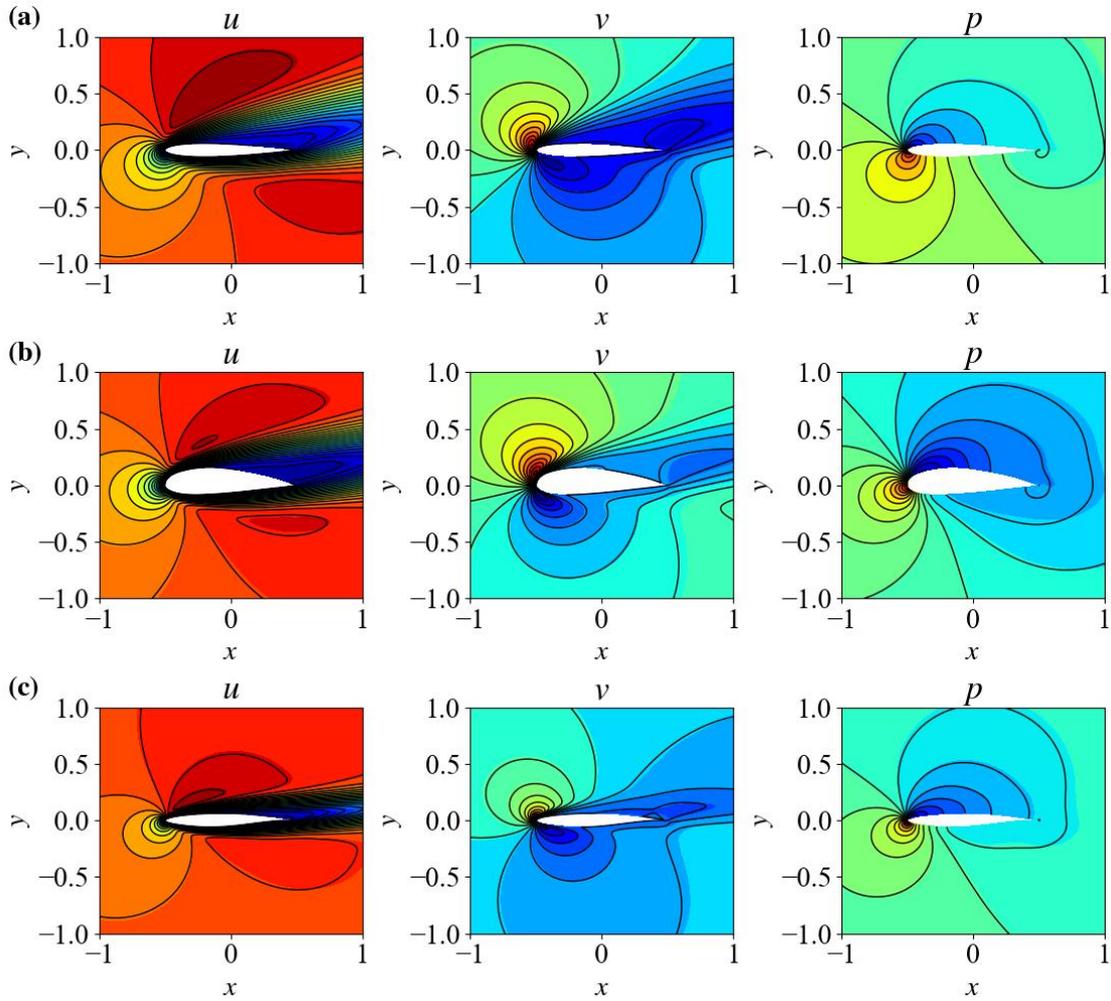

Figure 11. The contour plots of *u*, *v*, *p* predicted by Model 1 for (a) Case 1, (b) Case 2, and (c) Case 3. The solid black lines are the contour lines of the reference solution.

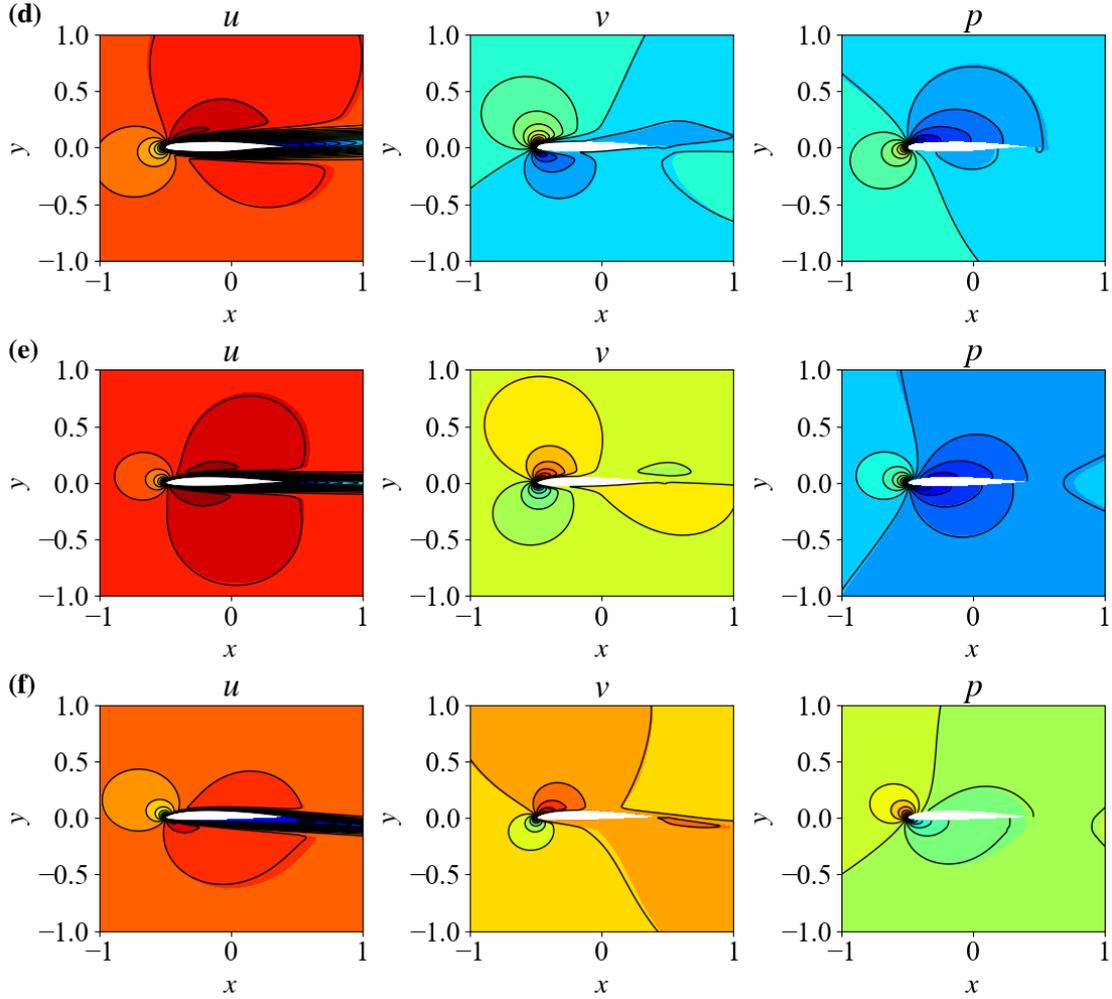

Figure 12. The contour plots of $u$, $v$, $p$ predicted by Model 1 for (d) Case 4, (e) Case 5, and (f) Case 6. The solid black lines are the contour lines of the reference solution.

Here, we further analyze the performance of three surrogate models in predicting the lift and drag curves of the NACA2412 airfoil at $Re$ = 500. Figure 13 illustrates the results, comparing the accuracy and generalization capabilities of the models. Within the training parameter space, both Model 1 and Model 2 achieve high prediction accuracy, closely matching the reference solutions. However, for test angles of attack outside the training parameter space, these models exhibit significant errors, highlighting their limited generalization to out-of-distribution samples. In contrast, Model 3, which incorporates a sufficiently large angle-of-attack parameter space during training, maintains excellent agreement with the reference solutions across the entire range of angles of attack, demonstrating superior generalization capability and robustness.

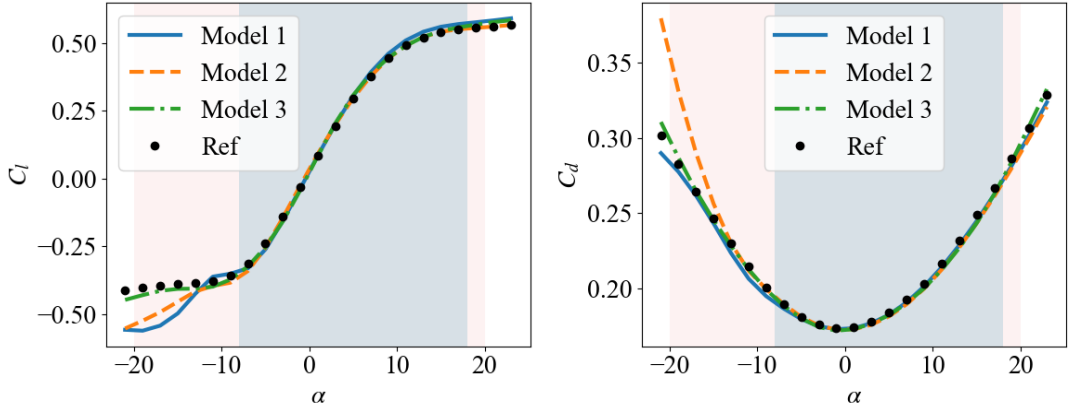

Figure 13. The lift and drag curves of the NACA2412 airfoil predicted by the three surrogate models. The blue region represents the parameter space used during the training of models 1 and 2, while the light coral region represents the parameter space used during the training of model 3.

## 4 Conclusions

In this study, we develop a parametric solver for laminar flows around airfoils based on time-stepping-oriented neural network and mesh transformation. It can solve a single flow as well as providing continuous solutions within a given parameter space. We first validate the effectiveness of the solver through several forward problem cases. The results demonstrate that the solver effectively and robustly handles laminar flows, achieving an average relative error of 3.6% for lift coefficients and 1.4% for drag coefficients.

Furthermore, we extend this solver to high-dimensional parametric engineering problems, covering nearly all laminar flows of interest around airfoils, and highlight that even simple fully connected networks exhibit strong generalization capabilities in various work conditions in engineering applications. The results show that using two different airfoil representation methods, $y$-coordinates and CST parameters, leads to problems with 16 or 204 dimensions, yielding similar convergence curves and accuracy. The solver efficiently solves all laminar flows within the parameter space in just 4.6 days, producing a surrogate model with an average lift coefficient error of 4.6% and an average drag coefficient error of 1.1%. More importantly, as demonstrated in the paper, the parameter space can be even flexibly expanded or reduced to meet specific requirements.

We emphasize that although TSONN does not outperform traditional numerical methods in accuracy or efficiency in solving single forward problem, its advantages in solving parametric problems deserve significant attention. It is expected that as TSONN's efficiency and accuracy in solving forward problems improve, it will show even greater potential in addressing parametric problems.

## Data Availability Statement

The data that support the findings of this study are available from the corresponding author upon reasonable request.

## Conflict of Interest Statement

The authors have no conflicts to disclose.

## Acknowledgments

We would like to acknowledge the support of the National Natural Science Foundation of China (No. 92152301).